# Triangular Spin-Orbit-Coupled Lattice with Strong Coulomb Correlations: Sn Atoms on a SiC(0001) Substrate

S. Glass[1], G. Li[2], F. Adler[1], J. Aulbach[1], A. Fleszar[2], R. Thomale[2], W. Hanke[2], R. Claessen[1], and J. Schäfer[1]

[1]*Physikalisches Institut and Röntgen Research Center for Complex Material Systems, Universität Würzburg, 97074 Würzburg, Germany*
[2]*Institut für Theoretische Physik und Astrophysik, Universität Würzburg, 97074 Würzburg, Germany*

Two-dimensional (2D) atom lattices provide model setups for Coulomb correlations inducing competing ground states, partly with topological character. Hexagonal SiC(0001) is an intriguing wide-gap substrate, spectroscopically separated from the overlayer and hence reduced screening. We report the first study of an artificial high-Z atom lattice on SiC(0001) by Sn adatoms, based on combined experimental realization and theoretical modeling. Density-functional theory of our √3-structure model closely reproduces the scanning tunneling microscopy. Instead of metallic behavior, photoemission data show a deeply gapped state (~2 eV gap). Based on our calculations including dynamic mean-field theory, we argue that this reflects a pronounced Mott insulating scenario. We also find indications that the system is susceptible to antiferromagnetic superstructures. Such spin-orbit-coupled correlated heavy atom lattices on SiC(0001) thus form a novel testbed for peculiar quantum states of matter, with potential bearing for spin liquids and topological Mott insulators.

*Design of high-Z atom lattices* – The wealth of physical phenomena in two-dimensional (2D) atom layers lends itself to both fundamental research and electronic applications. A key example is single-layer graphene, fabricated on a SiC(0001) substrate [1], which offers a high mobility attractive for use in devices. Intriguingly, graphene belongs to the family of 2D topological insulators that exhibit the quantum spin Hall effect [2] – yet lacks a sufficient energy gap that would allow room-temperature application. Therefore, growth of atom lattices with substantially higher *Z* number is a major research goal. Thereby, based on enhanced *spin-orbit coupling* (SOC), the bulk gap is expected to increase substantially. A key prediction is the Sn-based honeycomb lattice (stanene) with a gap magnitude of order 100 meV [3]. As a topological insulator (TI) it would offer spin-polarized edge channels for transport.

Of equal importance is the existence of *electronic correlations*: due to reduced screening in 2D, and boosted by substrates with a large band gap, the electrons will experience strong on-site repulsion. In several cases, where naively a half-filled metallic band of surface dangling bonds (DB) would be expected, this is known to lead to a Mott-Hubbard insulating state. Famous examples include the triangular surface reconstructions of the (√3×√3)-type formed on Si(111) and Ge(111) surface, by depositing 1/3 of a monolayer of Sn atoms [4,5]. Since the localized spins of the DBs must arrange with respect to each other, magnetic ordering at the surface is an expected consequence. Indeed, for Sn/Si(111), strong evidence argues for collinear antiferromagnetism (AFM) [6].

Restricting the choice of substrate to the semiconductors Si and Ge, however, is a limitation in conquering the rich phase diagram of systems as a function of *electron correlation* (manifest by Coulomb repulsion U, screening, hopping etc.) and the *strength of SOC* (favored for high Z). A rich plethora of phases, extending from spin-orbit coupled Mott insulators in general to specific scenarios such as topological Mott insulators, Weyl semimetals or spin liquids could be expected [7]. Therefore, extending the studies to a wide-gap substrate material is highly desirable. Here, SiC appears ideal, because its band gap of 3.2 eV (4H-SiC at 300 K) implies an almost five-fold increase over Ge (0.7 eV). As a result, the DB states are well separated away from the bulk bands, and one must expect a strong suppression of substrate screening, which favors strong Coulomb repulsion. Intriguingly, the chemically inert SiC may have a modified adatom bonding, which affects the hopping – but has not yet been widely explored.

A promising agenda thus is to place an atom array on top of SiC. To date, this has not been achieved experimentally, because a major obstacle is the poor surface smoothness as a result of its inertness against chemical treatments. Using hydrogen gas phase etching, the surface becomes tractable, and graphene can be grown [1]. However, a tremendous hurdle for use as a growth template for artificial adatoms (other than C) is the persistent H termination. Its thermal removal at ~800 °C usually results in a Si-induced (√3×√3)-reconstruction formed by mobile intrinsic Si atoms [8].

In this Letter, we report the first realization as well as experimental and theoretical characterization of an ordered Sn atom lattice on SiC(0001). Scanning tunneling microscopy (STM) of the triangular lattice in (√3×√3)-reconstruction is well reproduced by density-functional theory (DFT) in a structure model with Sn at $T_4$ sites. We find in photoemission that the system exhibits a large gap of ~2eV, and give arguments that the SiC substrate fosters this effect via large on-site Coulomb repulsion combined with suppressed electron hopping. We furthermore argue that a subtle interplay of magnetic and charge order appears likely in this new type of strongly correlated spin-orbit Mott insulator, which is a leap forward into the SOC-U phase diagram for surface systems.

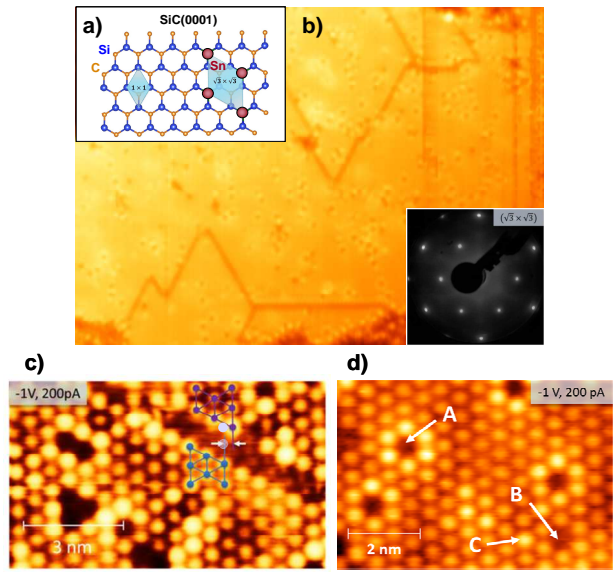

**Fig. 1. a)** Schematic of topmost buckled bilayer of the Si-face of SiC(0001), covered by a (√3×√3)-Sn reconstruction. **b)** LEED image (inset) and STM overview (-1.5 V, 100 pA, 60×42 nm) of (√3×√3)-Sn/SiC lattice. **c)** Close-up of registry shift with respect to the substrate. **d)** Characteristic point defects, as discussed in text.

*Technological approach* – The substrates were Si-face 4H-SiC(0001) with n-doping (conductivity ~0.05 Ohm·cm), and were subjected to chemical etching prior to loading into ultra-high vacuum. *In situ* the substrates have been etched in a H gas phase process, (details are described elsewhere [9]). This treatment removes the polishing damage, which cannot be achieved by wet-chemical methods alone. As a result the substrates exhibit large smooth terraces, which are H-passivated in a (1×1) pattern. Subsequent deposition of Sn was achieved in a novel *counterflow* approach, at a substrate temperature of ~750 °C. Thus, H gradually desorbs from the surface while, due to the immediate supply of Sn atoms, formation of the metal overlayer in the submonolayer regime can take place. A post-anneal for several minutes then leads to a well-ordered lattice in (√3×√3) reconstruction, observed in low-energy electron diffraction (LEED).

*Structure of the adatom lattice* – The hexagonal SiC(0001) surface is illustrated in **Fig. 1a)**. The surface layer is a buckled Si-C bilayer, which – for the substrate "Si-face" – carries Si atoms on top. For both the unreconstructed DB case as well as the H-passivated case, the surface has a (1×1) periodicity. The situation changes after adsorption of Sn on the H-free DB surface, where we experimentally find the (√3×√3) reconstruction. **Fig. 1b)** depicts the corresponding triangular LEED pattern and the large-area Sn coverage seen by STM. One also observes domain boundary lines and point defects, captured in close-up STM images **Fig. 1c)** and **1d)**, respectively.

The domain boundaries can be well explained as so-called registry shifts, where the (√3×√3) unit cell is shifted by one Si atom to the side. This leads to a trench-like fault line, with three possible directions (separated by 120°) on this hexagonal substrate. This is exactly what is observed in our experimental images **Fig. 1b)** and **1c)**. Characteristic point defects as in **Fig. 1d)** are in most cases surrounded by a ring of more intense appearing atoms (defect type A), while this is not always the case (type B), or even a single atom appears brighter (type C). All these defect patterns are well known from the "cousin" system (√3×√3)-Sn on Si(111), where their formation is discussed in terms of missing Sn atoms or an exchange with a substrate Si atom [4].

In order to establish a structural model for this system, we refer to similar systems formed by Sn and Pb atoms on the Si(111) and Ge(111) surfaces. For all these adatom-substrate combinations, (√3×√3)-reconstructions at 1/3 ML coverage have been reported. In particular, for Sn/Si(111) and Sn/Ge(111) the structure was probed by x-ray standing waves [10], or surface x-ray diffraction [11], respectively, and the Sn atoms were found to always reside on $T_4$ sites. Due to the close analogy with the current SiC(0001) system (regarding the $sp^3$-bonded bilayers) we conclude that here, too, Sn atoms are bonded to $T_4$ sites.

*Density-functional modeling* – To expose this structural model to a detailed test, we have performed DFT calculations. The Sn/4H-SiC(0001)-√3×√3 surface is modeled by a slab of three SiC bilayers and Sn adatoms with 1/3 monolayer coverage on the Si-face (the bottom C-face of the slab is H-saturated). For the simulation, the projector-augmented-wave method in the Vienna Ab Initio Simulation Package (VASP)

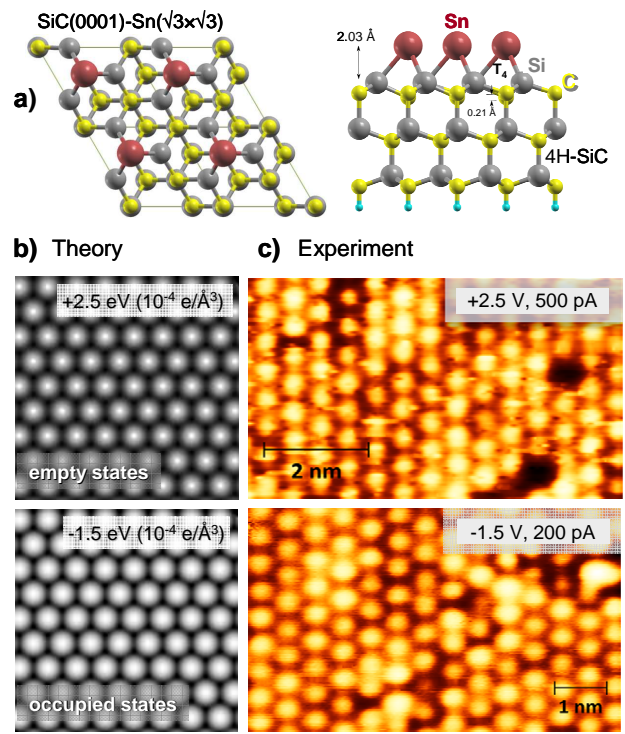

**Fig. 2** Structure model for DFT modeling, with Sn in $T_4$ sites on 4H-SiC(0001) (top and side view). **b)** Simulated STM images, and **c)** experimental STM images, for empty and occupied states, respectively. Close agreement is achieved for the whole range of tunneling parameters.

[12] within the local density approximation (LDA) was adopted (details in [13]). SOC was included as a second variational process. Sn is found to be energetically favored when staying at the $T_4$ position of the underlying Si layer, as illustrated in **Fig. 2a)**.

The STM images are simulated by using the Tersoff and Hamann approximation [14], where the details of the tip are neglected. This approximation leads to a simple expression of the tunneling current, involving only the local density of states (LDOS) of the surface. It can be visualized by generating iso-surfaces of the energy-integrated LDOS that mimics measurements at given bias and constant current, see **Fig. 2b)** for negative bias (occupied states) and positive bias (empty states), respectively. Regardless of the bias, however, the pattern seen is of triangular character, without any further substructure.

An experimental STM bias series can then serve as critical scrutiny for the structural model, and as a test of the predicted electronic properties. From a range of bias values, in **Fig. 2c)** we have compiled close-up images at -1.5V (occupied states) and +2.5 V (unoccupied states). For all bias values, we find a triangular pattern with circular elevations (allowing for small deviations ascribed to the tip). Importantly, we never observe a change of the pattern that would result from a different structure (e.g., hexagonal rings or a substructure) that would reflect trimers instead of monomers at each site.

*Electronic properties* – The DFT approach outlined above also provides the electronic band structure including SOC, which is plotted in **Fig. 3a)**. The calculations predict a metallic ground state with a half-filled narrow DB band within a large band gap, which is separated from the valence band by ~1 eV and from the conduction band by ~1.7 eV. Two further occupied surface bands are found in the 1.2 – 1.4 eV region below $E_F$. The metallic band, enlarged in **Fig. 3b)**, is characterized by i) a substantial SOC splitting, with maxima shifted in momentum away from the Γ-point, and ii) a very small band width of only ~ 0.29 eV. This is notably much smaller than the DFT band width for Sn on Si(111), which is ~0.50 eV [6]. The Sn/SiC surface band mainly consists of the Sn $p_z$-orbitals, with the corresponding Wannier function extending down into the first SiC bilayer, indicating a hybridization of the adatom Sn states with those of the SiC substrate.

The experimental *spectral function* has been determined by ultraviolet k-integrated photoemission the results of which are shown in **Fig. 4a)**. Most importantly, one observes strong *suppression of spectral weight* at the chemical potential, and a peak of the signal at about ~1 eV below. Another upswing of intensity at about 2.5 eV below $E_F$ must be assigned to a combination of the second (filled) surface band and the substrate bands.

Owing to the Sn adatoms being situated atop a wide-gap (low screening) substrate, on-site Coulomb correlation U will play a major role. This typically leads to a Mott insulating state, where the electron removal and addition spectra (i.e., lower and upper Hubbard band) are separated by a gap comparable to U, as in **Fig. 4b)**. Thus, the actual correlated DOS varies substantially from the DFT picture, where – per definition – such effects are not included.

The photoemission spectral features clearly argue for a Mott insulating state. They bear close resemblance to what has been observed for √3-Sn on Si(111) [6]. However, there the peak of the lower Hubbard band is located only ~0.4 eV below $E_F$, while

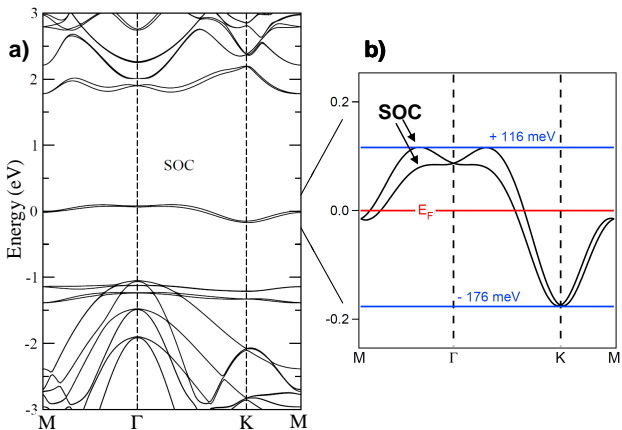

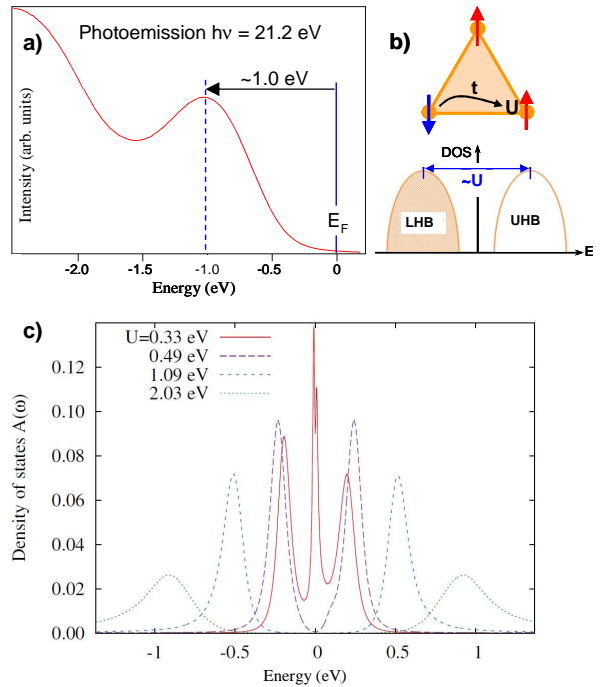

**Fig. 3 a)** Band structure of √3-Sn on 4H-SiC(0001) from DFT including SOC. **b)** Close-up of the DB band at the Fermi level, which (without U) is metallic and has a very narrow bandwidth of only ~0.29 eV. SOC induces a notable band splitting.

**Fig. 4 a)** Experimental spectral function obtained from photoemission (He-I excitation, 300 K). The system is in an insulating state with a half-gap of ~1.0 eV. **b)** Coulomb correlations (U) with hopping (t) split the metallic DB state into lower and upper Hubbard bands (LHB, UHB). **c)** Spectral function from DMFT, requiring high U values to approximate the experimental spectrum.

in the current system it is much lower, namely ~ 1eV, which implies a U value of ~2 eV.

The only other adatom system known on SiC(0001) is the much less controlled auto-formation of a √3-reconstruction by (low-Z) Si via mobile substrate atoms [8]. Photoemission data also suggest a Mott-insulating case [15,16], and U values ranging from 1.6 – 2.1 eV have been inferred [17,18].

*Many-body aspects: hopping and screening –* Since the in-plane lattice constant of SiC(0001) is 20% smaller than for Si(111), the adatom spacing is reduced, and one would therefore expect that both hopping parameter and resulting bandwidth of the DB state are increased for SiC(0001). Contrary to this intuition, the opposite is the case: the LDA bandwidths amount to ~0.50 eV in Sn/Si(111), yet only ~0.29 eV in Sn/SiC(0001). This unexpected and significant reduction indicates that the hopping process is not decoupled from the substrate. Rather, an indirect hopping path through it must be considered, as emphasized recently also by Lee *et al.* [19].

The fact that "indirect hopping" across SiC is less efficient can be connected to the higher "chemical inertia" compared to Si and Ge. SiC is a partly ionic material (due to different electronegativities of Si and C), where hybridization between adatom and substrate orbitals will be much reduced. This results in small effective hopping across the substrate. In a single-band tight-binding model for the DB, we find that the nearest-neighbor hopping amplitude reduces from 52.7 meV for Sn on Si(111) [20] to 27.3 meV for the same atoms on SiC(0001).

Furthermore, SiC as wide-gap insulator (gap 3.2 eV, while 1.1 eV in Si) is characterized by a substantially less effective dielectric screening, which will contribute to a larger U at the Sn adatom. For dielectric screening only the electronic contribution is relevant, while ionic vibrations are negligible in dynamic screening. Regarding the nonpolar Si, both (static and electronic) dielectric constants have the same value of ~11.7. For SiC, the static dielectric constant, which involves the ionic charge, amounts to $\varepsilon_S$(SiC) ~9.7, while the electronic (high-frequency) dielectric constant is only $\varepsilon_\infty$(SiC) ~6.5 [21]. This implies a reduction of the screening in SiC by a factor of ~2.

Thus, the outstanding characteristics of the SiC substrate are twofold: (i) it makes the hopping small and therefore the DB band flat, and (ii) the small dielectric constant allows only a limited screening of the local Coulomb interaction (Hubbard U) on the adatoms. Together this leads to a considerable strengthening of electronic correlations.

For the present Sn/SiC system, we examined the electronic correlations by a LDA + dynamic mean-field theory (DMFT) study [22], where the constructed tight-binding model is subject to an on-site Coulomb repulsion. **Fig. 4c)** shows the density of state (DOS) of this surface band. As one can see, the intensity at the chemical potential (ω = 0) is gradually lost with the increase of U, signaling a metal-insulator transition. Thus, strong Coulomb repulsion can certainly be a key ingredient to drive the system to an insulating ground state. For a correlation value of U ~ 2 eV we find reasonable qualitative agreement with the photoemission behavior, i.e., the peak for the lower Hubbard band is shifted to ~U/2 = 1 eV below $E_F$.

*Long-range interactions and magnetic ordering –* In addition to electronic correlations, further driving forces which favor symmetry breaking can come into play. We find that a characteristic feature shared by Sn/Si(111) and Sn/SiC(0001) is the presence of strong next-nearest-neighbor (NNN) hopping in the DB tight-binding description. In such a single-band model, the substrate states are projected out, and the hybridization of the adatom states with those of the substrate is translated into an *effective hopping* of the adatom lattice. The presence of strong NNN hopping in the effective tight-binding model was found vital to explaining the experimental evidence, as it drives the magnetic ordering of the Sn/Si(111) system [6] from 120°-AFM to collinear AFM in the presence of electronic interactions with U ~ 0.7 eV.

The role of NNN hopping has recently received independent support from a DFT study (with hybrid exchange-correlation functional to address the self-interaction error) in large supercells [19], pointing out that the substrate interaction in Sn/Si(111) favors a collinear AFM state. In this spirit, for Sn/SiC we compared the total energy of various magnetic configurations. Spin-polarized calculations carried out in a (3×6) unit cell are compatible with all four magnetic structures, i.e., ferromagnetism, ferrimagnetism, 120°-AFM, and collinear AFM (no SOC and electronic interaction beyond LDA was considered). The results in Table I show that, similar to Sn/Si(111), collinear AFM is the most stable configuration.

**Table I:** Total energy (in units of meV per (√3×√3) unit cell) of four different magnetic structures relative to the non-magnetic ground state.

| Magnetic Configuration | Total energy per unit cell |
|---|---|
| Ferromagnetism | + 3.03 meV |
| Ferrimagnetism | – 2.03 meV |
| 120° - AFM | – 2.75 meV |
| Collinear AFM | – 2.93 meV |

Notably, since collinear AFM, 120°-AFM, and ferrimagnetism are nearly energy-degenerate, it is not clear whether the real-world system will condense into one specific ordered state. This leads to the exciting question whether a *spin-liquid* may be encountered, with rare experimental evidence to date [23]. By definition, however, that approach alone (based on DFT), which neglects on-site electron correlations, cannot reproduce the broadened many-body spectral function observed in photoemission. In particular, in our calculation of the DOS for collinear AFM, the energy gap between occupied and empty states amounts to only ~0.1 eV, in stark contrast to the ~2 eV deduced from the photoemission experiment. This discrepancy further underpins the role of strong Coulomb correlations, as in our LDA+DMFT results. Interestingly, they may serve to enhance the stability of collinear AFM, as in Sn/Si(111) [5], which deserves further careful investigation.

*Competing orders and doping* – However, a competing order may play a role in the Sn/SiC system, i.e. charge ordering (CO). Nonlocal interactions (i.e., charge fluctuations) in triangular lattices on Si(111) were found by GW+DMFT to give rise to screening corrections, and it was argued that (3×3) CO may set in [24]. Its characteristic (3×3) superstructure, however, can be experimentally distinguished from that resulting from collinear AFM with its threefold rotational degenerate (2√3×2√3) pattern [6]. A second individual fingerprint is the spectral weight transfer from band backfolding. Charge and spin order on the triangular lattice may even coexist [25], calling for a critical test on SiC.

Doping of a Sn layer by substitutional atoms was achieved for Si(111) [26], yet impurities can give rise to unwanted scattering. An alternative may emerge with SiC, since a wide-gap material has less intrinsic carriers at RT. Backside electrostatic gating can then tune the chemical potential in the surface layer at will. For doped triangular lattices, a most exciting question is the occurrence of unconventional superconductivity (SC). Surface SC was reported for (√7×√3) reconstructions of Pb and In on Si(111) [27]. Therefore, it is still an outstanding challenge to generate SC in a surface-based triangular lattice. On theoretical grounds, unconventional SC with chiral (time-reversal symmetry-breaking) d+id pairing should exist [28,29].

*Interplay of SOC and correlations* – An important ingredient of the current Sn/SiC system is the substantial SOC. It has just recently been shown that in bulk α-Sn the SOC suffices to drive the formation of a TI [30]. It is even conceivable to further enhance the SOC by using Pb or Bi in our epitaxial process on SiC(0001). Moreover, it has been demonstrated theoretically that a 2D triangular surface lattice can host a TI state [31]. With regard to the strong correlations, the existence of *Topological Mott Insulators*, where interactions trigger the band inversion, was put forward for the honeycomb lattice [32]. Our realization of a triangular lattice with both SOC and strong correlations on SiC(0001) may be another candidate. Also, we point out that the lattice constant of the theoretically proposed hexagonal TI stanene [3] and SiC(0001) are very close. Therefore, our work strongly encourages further exploration of Sn-based lattices on SiC.

This work was supported by the Deutsche Forschungsgemeinschaft (DFG) within FOR 1162. We acknowledge the Jülich Supercomputing Centre for providing the computer resources. G.L., W.H., and R.T. have been supported by the European Research Council through ERC-StG-Topolectrics-336012.